\documentclass[aps,prd,showpacs,amssymb]{revtex4}
\usepackage{graphicx}% Include figure files
\usepackage{bm}% bold math

\begin{document}
%\LARGE
%\preprint{gr-qc/}
%\draft
\title{Semiclassical back reaction around a cosmic dislocation}
\author{V. A. De Lorenci}
 \email{delorenci@unifei.edu.br}
\author{R. Klippert}
 \email{klippert@unifei.edu.br}
\author{E. S. Moreira Jr.}
 \email{moreira@unifei.edu.br}
\affiliation{Instituto de Ci\^encias Exatas,
Universidade Federal de Itajub\'a, 
Av.\ BPS 1303 Pinheirinho, 37500-903 Itajub\'a, MG, Brazil} 

\date{September 27, 2004}
%\twocolumns[
%\hsize\textwidth\columnwidth\hsize\csname@twocolumnfalse\endcsname 

\begin{abstract}

The energy-momentum vacuum average of a conformally coupled 
massless scalar field vibrating 
around a cosmic dislocation ---
a cosmic string with a dislocation along its axis ---
is taken as source of the linearized semiclassical 
Einstein equations. The solution up to first 
order in the Planck constant is derived. 
Motion of a test particle is then discussed, showing that
under certain circumstances a helical-like dragging effect,
with no classical analogue around the cosmic dislocation,
is induced by back reaction.

\end{abstract}

%]
\maketitle

%\begin{multicols}{2}

%
%\section{Introduction}
%\hspace{0.5cm} 
%
\section{Introduction}
Various aspects of fields 
around cosmic strings have been considered  
in the literature \cite{vil94,bor01}. 
The usual motivation for such investigations is that
these objects may play some role in the cosmological scenario.
There is though another equally appealing motivation. Namely,
gravitational fields of cosmic strings are locally flat solutions of
the Einstein equations, resulting that one can carry out calculations
without meeting serious obstacles, and still revealing non trivial effects.
Moreover, the physical content of such effects may not only be 
relevant in the context of classical and
semiclassical general relativity, but also 
in condensed matter physics,  where a 
geometric interpretation can be used to describe
physical properties of some linear defects in solids 
\cite{kro81}.

As is well known, there is no Newtonian potential around an ordinary cosmic
string  of mass density $\mu$ \cite{vil94}. 
Accordingly, a particle left at rest near such object 
will remain at rest.
The study of semiclassical back reaction on the gravitational field 
of an ordinary cosmic string
has shown that vacuum fluctuations (of a conformally coupled massless scalar field)
induce a Newtonian potential ($c=1$, unless stated otherwise),
\begin{equation}
\Phi(\rho)=-\frac{G\hbar\ {\cal F}(\mu)}{\rho^{2}},
\label{npotential}
\end{equation}
at a distance $\rho$ from the cosmic string \cite{his87,ban91,cam95,ali98}. 
Subjected to $\Phi(\rho)$, a particle
initially at rest experiences a presumably small, but non vanishing 
``quantum mechanical" force when $\mu\neq 0$.

The gravitational field of an ordinary cosmic string corresponds to the geometry
of a conical spacetime, where the deficit angle is proportional to $\mu$ \cite{vil94}. 
It has been conjectured in Ref. \cite{gal93} that the gravitational field
of a certain type of chiral cosmic string \cite{bek92} may correspond to the
geometry of a cosmic dislocation --- a conical spacetime with a helical structure.
It seems pertinent to extend investigations on semiclassical
back reaction to this background, and that is the issue addressed here.

In Sec. II, the geometry of a cosmic dislocation is presented. 
In the following section,
the vacuum average of the energy-momentum tensor 
for a conformally coupled massless scalar field \cite{mor03} is used
as source of the Einstein equations. The solution
up to first order in $\hbar$ is determined, and then used in Sec. IV to discuss
the induced force acting on a test particle. Sec. V closes the work 
with a summary and additional remarks.

\section{The classical background}

A cosmic dislocation has as line element \cite{gal93,tod94}
\begin{equation}
ds^2=dt^2 - dr^2 - \alpha^2 r^2 d\theta^2 - (dz + \kappa\  d\theta)^2,
\label{cdle}
\end{equation}
with the usual identification 
$(t,r,\theta,z) \sim (t,r,\theta +2\pi,z)$.
The geometry in Eq. (\ref{cdle}) is obtained from that of a conical spacetime
with deficit angle $2\pi(1-\alpha)$, replacing $dz$ by $dz+\kappa\ d\theta$.
When the cone parameter $\alpha$
equals unity and the dislocation parameter $\kappa$ vanishes, Eq. (\ref{cdle})
becomes the Minkowski line element written in circular cylindrical coordinates.

In fact, by defining new coordinates 
$\varphi := \alpha \theta $ and $Z := z + \kappa \theta$, 
Eq. (\ref{cdle}) becomes
\begin{equation}
ds^{2}=dt^{2}-dr^2 -r^2 d\varphi^2 - dZ^2,
\label{fle}
\end{equation}
where the identification
\begin{equation}
(t,r,\varphi,Z) \sim (t,r,\varphi+2\pi\alpha,Z+2\pi\kappa)
\label{bcondition}
\end{equation}
must be observed. It is clear from Eq. (\ref{fle})
the locally flat nature of the background, 
and from Eq. (\ref{bcondition}) its helical structure.
A particle at rest will
remain at rest and (locally) geodesics are simply straight lines.

It should be remarked that the geometry  in Eq. (\ref{cdle}) fits in
general relativity \cite{gal93} as well as in the Einstein-Cartan theory \cite{tod94}.
In the context of general relativity there is a curvature singularity along the symmetry
axis. In the Einstein-Cartan theory, there is also a torsion singularity along the
symmetry axis, when $\kappa\neq 0$.

\section{Back reaction}

The non trivial global geometry encoded in Eq. (\ref{bcondition}) leads to vacuum
polarization. Although vacuum averages of the fields themselves
vanish, that is not necessarily the case for their renormalized energy-momentum tensors 
$\left< T^{\mu}{}_{\nu} \right>$. 
In particular, for a conformally coupled  massless scalar field, the behavior
of $\left< T^{\mu}{}_{\nu} \right>$ as $\kappa/r\ll 1$
[in terms of the coordinates in Eq. (\ref{fle})] is given by  \cite{mor03}
\begin{equation}
\left< T^{\mu}{}_{\nu} \right> =\frac{\hbar}{r^4}
\left(
     \begin{array}{cccc}
      -A &  0 & 0                    & 0 \\
       0 & -A & 0                    & 0 \\
       0 &  0 & 3A                   & \kappa B/r^2 \\
       0 &  0 & \kappa B             & -A
     \end{array}
\right),
\label{emtensor}
\end{equation}
where $A(\alpha):=(\alpha^{-4}-1)/1440\pi^2$ and 
\begin{widetext}
\begin{eqnarray}
B(\alpha): = \frac{1}{32\pi^3\alpha^2}\int_0^\infty\! d\tau\
\frac{\alpha\sin(\pi/\alpha)\,
[\, \cos(\pi/\alpha) -\cosh(\tau)+
\tau\sinh(\tau)\, ] - 
\pi[\cos(\pi/\alpha)\cosh(\tau)-1]}
{[\,\cosh(\tau)-\cos(\pi/\alpha)\,]^2\cosh^4(\alpha\tau/2)}.
\label{b}
\end{eqnarray}
\end{widetext}
Noting that $A(1)=0$ and 
\begin{equation}
B(1)=\frac{1}{60\pi^{2}}, 
\label{nfactor}
\end{equation}
when $\alpha=1$ 
and $\kappa=0$ it follows that 
$\left< T^{\mu}{}_{\nu} \right>= 0$, 
which is consistent with the fact that the vacuum averages in
Ref. \cite{mor03} are renormalized with respect to the 
Minkowski vacuum. [It should be pointed out that Eq. (\ref{emtensor})
holds for $\kappa\neq 0$ only when $\alpha\neq 1$, since for 
$\alpha=1$ the diagonal components vanish and subleading contributions
depending on $\kappa$ should be taken into account.]

The semiclassical approach to back reaction in general relativity 
is implemented by feeding the Einstein equations with 
the vacuum average in Eq. (\ref{emtensor}).
Since $\left< T^{\mu}{}_{\nu} \right>$ is traceless, these equations are
\begin{equation}
R^{\mu}{}_{\nu}=-8\pi G\left< T^{\mu}{}_{\nu} \right>.
\label{einstein}
\end{equation}
The most general stationary axially symmetric line element, 
symmetric also with respect to translations along the axis, is 
\begin{equation}
ds^{2}=g_{00}(r)dt^{2}+2g_{02}(r)dt\ d\varphi
+g_{11}(r)dr^{2}
+g_{22}(r)d\varphi^{2}+2g_{23}(r)d\varphi\ dZ+g_{33}(r)dZ^{2},
\label{sle}
\end{equation}
whose form is clearly invariant under redefinition of the
radial coordinate $r\rightarrow\rho$, $r=f(\rho)$. 
This gauge freedom can be used to choose 
$g_{22}(r)=-r^{2}$.
The next step is to allow quantum perturbations of Eq. (\ref{fle}) 
consistent with Eq. (\ref{sle}). 
Then, discarding non physical solutions, the standard
linearization procedure \cite{lan51,his87,cam95,ali98} 
applied to Eqs. (\ref{einstein}) leads to
\begin{widetext}
\begin{eqnarray}
ds^2 &=& \left(1-\frac{4\pi AG\hbar }{r^2}\right)(dt^2-dZ^2) 
-\left(1+\frac{16\pi AG\hbar}{r^2}\right)dr^2
-r^2 d\varphi^2 -\frac{4\pi\kappa B G\hbar}
{r^2} \ d\varphi\ dZ,
\label{breaction}
\end{eqnarray}
\end{widetext}
up to first order in $\hbar$. 
When $\kappa\neq 0$ and/or $\alpha\neq 1$,
an inspection shows
that the perturbed background is locally flat only asymptotically 
(i.e., as $r\rightarrow\infty$), and that the dislocation contributes 
($\kappa\neq 0$)
with non vanishing off-diagonal components in the semiclassical metric tensor.

When $\kappa=0$, Eq. (\ref{breaction}) should reproduce the results
in the literature regarding semiclassical 
back reaction on the gravitational field 
of an ordinary cosmic string. In order to implement this check, 
one uses the gauge freedom in choosing the radial coordinate, 
defining
\begin{equation}
\rho:=r+\frac{\lambda\pi A G\hbar}{r},
\label{newr}
\end{equation}
where $\lambda$ is an arbitrary dimensionless parameter.
Thus, Eq. (\ref{breaction}) becomes
\begin{widetext}
\begin{eqnarray}
ds^2 &=& \left(1-\frac{4\pi AG\hbar }{\rho^2}\right)(dt^2-dZ^2) 
-\left(1+(\lambda+8)\frac{2\pi AG\hbar}{\rho^2}\right)d\rho^2
-\rho^2\left(1- \lambda\frac{2\pi AG\hbar}{\rho^2}\right)d\varphi^2 
\nonumber
\\
&&-\frac{4\pi\kappa B G\hbar}
{\rho^2}\  d\varphi\ dZ.
\label{nbreaction}
\end{eqnarray}
\end{widetext}
For $\kappa=0$, the results in Refs.  
\cite{his87} and \cite{ali98} are obtained from 
Eq. (\ref{nbreaction})
by setting $\lambda=-10$ and $\lambda=-4$, respectively.

\section{Dragging effect}

A physical consequence of the perturbed geometry in Eq. (\ref{nbreaction})
is revealed when studying the geodesic motion. In order to get rid of ``inertial
forces", one sets $\lambda=2$ in Eq. (\ref{nbreaction}), resulting
\begin{widetext}
\begin{eqnarray}
ds^2 &=& \left(1-\frac{4\pi AG\hbar }{\rho^2}\right)
(dt^2-\rho^{2}d\varphi^2-dZ^2) 
-\left(1+\frac{20\pi AG\hbar}{\rho^2}\right)d\rho^2
-\frac{4\pi\kappa B G\hbar}
{\rho^2}\  d\varphi\ dZ.
\label{iframe}
\end{eqnarray}
\end{widetext}
Using the coordinate time $t$ (instead of the proper time) as
an affine parameter, the geodesic equations corresponding to this gauge, 
up to first order in $\hbar$, yield (inserting dimensionful $c$)
\begin{widetext}
\begin{eqnarray}
&&a_{\rho}=\frac{4\pi G\hbar}{c\rho^{3}}
\left[-A\left(1-7\frac{v_{\rho}^{2}}{c^{2}}+
5\frac{v_{\varphi}^{2}}{c^{2}}-\frac{v_{Z}^{2}}{c^{2}} \right) 
-\frac{\kappa B}{\rho}\frac{v_{\varphi}v_{Z}}{c^{2}}\right],
\label{rforce}
\\
&&a_{\varphi}=\frac{4\pi G\hbar\kappa B}{c\rho^{4}}\frac{v_{\rho}v_{Z}}{c^{2}},
\label{aforce}
\\
&&a_{Z}=\frac{8\pi G\hbar\kappa B}{c\rho^{4}}\frac{v_{\rho}v_{\varphi}}{c^{2}},
\label{axforce}
\end{eqnarray}
\end{widetext}
where $(v_{\rho},v_{\varphi},v_{Z})$ and  $(a_{\rho},a_{\varphi},a_{Z})$
are the usual circular cylindrical components of the particle velocity and
acceleration in Euclidean space (i.e.,
$v_{\rho}=\dot{\rho}$,
$v_{\varphi}=\rho\dot{\varphi}$,
$v_{Z}=\dot{Z}$, 
and 
$a_{\rho}=\ddot{\rho}-\rho\dot{\varphi}^{2}$,
$a_{\varphi}=\rho\ddot{\varphi}+2\dot{\rho}\dot{\varphi}$,
$a_{Z}=\ddot{Z}$, with the overdot denoting differentiation 
with respect to the coordinate time $t$). 
In order to interpret Eqs. (\ref{rforce}) to (\ref{axforce})
in the context of a possibly realistic scenario, one should recall that
the physics of formation of ordinary cosmic strings
suggests $\alpha<1$ and very close to unity \cite{vil94}, 
in which case Eq. (\ref{nfactor})
is a good approximation.

As the quantum corrections in Eq. (\ref{iframe}) are ``small",
the corresponding frame is nearly inertial. The right-hand sides
of Eqs. (\ref{rforce}) to (\ref{axforce}), 
after multiplied by the mass of a particle,
can be interpreted as the components of a 
gravitational force acting on the particle. 
By setting $\kappa=0$, the last term in Eq. (\ref{rforce})
as well as $a_{\varphi}$ and $a_{Z}$ vanish. 
Additionally, if the velocity is much smaller than $c$, 
$a_{\rho}=-4\pi G\hbar A/c\rho^{3}$, which matches the Newtonian potential in
Eq. (\ref{npotential}).

The contributions 
due to the dislocation ($\kappa\neq 0$) 
in the equations of motion (\ref{rforce}) to (\ref{axforce})
are non vanishing, only if 
the corresponding transverse velocities are both non vanishing. 
If the particle is initially moving on a plane defined by 
$v_{\rho}=0$
with $v_{\varphi}v_{Z}>0$, 
then $a_{\varphi}=a_{Z}=0$, 
whereas the last term in Eq. (\ref{rforce}) 
corresponds to an attractive or repulsive force, 
for $\kappa>0$ or $\kappa<0$, respectively.
If the initial motion takes place on a plane defined by
$v_{\varphi}=0 $ 
with $v_{\rho}v_{Z}>0$,
$a_{Z}$
and the last term in $a_{\rho}$ vanish, and the force corresponding to 
$a_{\varphi}\neq 0$ sweeps the particle
around the symmetry axis, counterclockwise or clockwise, for
$\kappa>0$ or $\kappa<0$, respectively.
Finally, if the particle is initially moving on a plane defined by
$v_{Z}=0 $
with $v_{\rho}v_{\varphi}>0$, 
$a_{\varphi}$
and the last term in $a_{\rho}$ vanish, 
whereas the force corresponding to $a_{Z}\neq 0$
pushes the particle up or down,
for $\kappa>0$ or $\kappa<0$, respectively
[this is the only effect which has a classical analogue, 
as the equations of motion are recast in terms of the coordinates
$(t,r,\theta,z)$].

In considering the  ``dragging" effect described above, 
a pertinent issue that might be raised concerns the
dependence of it on the choice of a particular radial coordinate
[such as the one corresponding to $\lambda=2$ in Eq. (\ref{newr})].
This effect amounts to state that, for
$\dot{\varphi}=0$ and non vanishing transverse velocities,
$\ddot{\varphi}\neq 0$ when $\kappa\neq 0$.
By observing Eq. (\ref{aforce})
one sees that redefinition of 
the radial coordinate $r=f(\rho)$ does not change this fact.
In other words, the dragging effect is gauge invariant.

\section{Final remarks}

In summary, this work extended investigations on semiclassical
general relativity in ordinary conical spacetime to a cosmic dislocation,
whose geometry has been conjectured to be associated with the gravitational
field of a certain chiral cosmic string.
By perturbing the locally flat metric tensor,
the main result is that back reaction due to a conformally coupled
massless scalar field leads
to non vanishing off-diagonal components in 
the semiclassical metric tensor,
resulting in a helical-like dragging effect 
on the motion of test particles.
As the dragging effect has no classical counterpart
around the cosmic dislocation, it might (in principle)
play a role in astrophysical or cosmological scenarios.

Before closing, it should be mentioned that semiclassical dragging effects 
around cosmic strings were first suggested in Ref \cite{mat90}, in the
context of vacuum polarization around a spinning cosmic string
(a cosmic string with a time dislocation \cite{maz86,gal93,tod94}).
The associated background, however, is not globally hyperbolic leading
to pathological vacuum fluctuations \cite{lor01}.

\begin{acknowledgments}

This work was partially supported by the 
Brazilian research agencies CNPq and FAPEMIG.

\end{acknowledgments}

%\end{multicols}

\end{document}